\documentclass[a4paper,prb,amsmath,amssymb,twocolumn]{revtex4}
\usepackage[dvips]{graphicx}
\usepackage{psfrag}
\newcommand{\beq}{\begin{equation}}
\newcommand{\eeq}{\end{equation}}

\begin{document}
\author{Christian Joas, J\"urgen Dietel, and Felix von Oppen}
\affiliation{Institut f\"ur Theoretische Physik, Freie Universit\"at
  Berlin, Arnimallee 14, D-14195 Berlin, Germany} 
\title{Microwave photoconductivity of a two-dimensional electron gas
  due to intra-Landau-level transitions } 
\date{\today}
\begin{abstract}
  Motivated by a recent experiment, we study the microwave-induced
  photoconductivity of a two-dimensional electron gas arising from
  intra-Landau-level transitions within a model
  where the electrons are
  subject to a unidirectional periodic potential in addition to a
  weaker impurity potential.  
  With appropriate identifications, our results can be compared to 
  experiment and allow us to explain the 
  sign of the photocurrent, its dependence on magnetic field and 
  microwave frequency
  as well as the microwave-induced suppression of the 
  Shubnikov-deHaas oscillations. 
\end{abstract}
\maketitle

\section{Introduction}
Recently, novel magnetooscillations of the microwave photoconductivity
were discovered in ultra-high-mobility two-dimensional (2d) electron
systems in the presence of a weak magnetic field.
\cite{Mani,Zudov,Yang, Dorozhkin,Willett} For sufficiently strong
microwave intensity, the longitudinal photoconductivity can become
close to zero in certain magnetic field regions, with seemingly
activated temperature dependence. Specifically, such so-called
``zero-resistance states" occur whenever the microwave frequency
$\omega$ is related to the cyclotron frequency $\omega_c$ as $\omega =
(k + \alpha) \omega_c$ ($k=1,2,3 \ldots$; $\alpha$ is a constant phase
shift).

Several mechanisms have been proposed to account for the photocurrent
oscillations.
\cite{Andreev,Bergeret,Durst,Shi,Lei,Vavilov,Ryzhii,Dmitriev1,Dmitriev2,Koulakov,Cooper,Dietel04,Auerbach}
One mechanism is based on the observation that disorder-assisted
microwave absorption is accompanied by a real-space displacement which
depending on the magnetic field, is preferentially along or against
the applied dc electric field.\cite{Durst,Shi,Lei,Vavilov,Ryzhii} We
refer to this mechanism as {\it displacement mechanism} (DP). A second
contribution to the photoconductivity arises from the
microwave-induced change in the electronic distribution
function.\cite{Dorozhkin,Dmitriev1,Dmitriev2} It turns out that
typically, this {\it distribution-function mechanism} (DF) tends to
dominate the magnetooscillations of the photoconductivity in realistic
samples,\cite{Dmitriev2} although there are exceptions to
this.\cite{Dietel04} Zero-resistance states are expected to occur once
the microwave-induced oscillations become so strong that the {\it
  microscopic} longitudinal conductivity is negative within certain
magnetic-field regions.\cite{Andreev,Bergeret}

In a very recent experiment, Dorozhkin {\it et al.}\cite{Dorozhkin04}
focused on the regime $\omega\ll\omega_c$. Unlike in previous
experiments, the microwave irradiation can then only induce {\it
  intra}-LL transitions.  According to Dorozhkin {\it et al.}, a
considerable microwave-induced reduction of the diagonal conductivity
is observed in this regime, but it appears that no zero-resistance
states were found. This overall reduction of the diagonal conductivity
is accompanied by a significant suppression of the Shubnikov-deHaas 
oscillations. 

In this paper, we investigate the effect of microwave-induced
intra-Landau-level transitions on the photoconductivity within a model
in which the 2DEG is subjected to a unidirectional and {\it static}
periodic potential. As shown in Ref.\ \onlinecite{Dietel04}, this
model allows one to compute the photoconductivity using Fermi's golden
rule which, in the appropriate geometry, leads to results which are
parametrically consistent with those for disorder-broadened Landau
levels. We find that our results are consistent with the principal
experimental findings and predict a periodic-potential induced
anisotropy of the intra-LL photocurrent.

This paper is organized as follows. In Sec.\ \ref{model} we introduce
the model and discuss the basic processes which contribute to the
photoconductivity for $\omega\ll\omega_c$. In Sec.\ \ref{calc} , we
explicitly compute the photoconductivity in this regime, including
both the displacement and the distribution-function mechanism.  Our
results are compared to experiment in Sec.\ \ref{exp} and summarized
in Sec.\ \ref{concl}.

\section{Model and basic processes}
\label{model}

We consider a two-dimensional electron gas (2DEG) subjected to a
perpendicular magnetic field $B$ and a unidirectional, static
modulation potential $V({\bf r})=V\cos\left(Qx\right)$ of period
$a=2\pi/Q$. We assume that the modulation potential $V({\bf r})$
exceeds the residual disorder potential $U({\bf r})$, whose correlator
$W({\bf r})$ falls off isotropically on the scale of the correlation
length $\xi$. As appropriate for a high-mobility 2DEG, we assume a
smooth disorder potential with correlation length $\xi\gg \lambda_F$.
($\lambda_F$ denotes the zero-field Fermi wavelength.) 

The 2DEG is irradiated by microwaves described by the electric
potential
\begin{equation}
\phi({\bf r},t) = -{\frac{e}{2}} {\bf r} ({\bf E}^*e^{i\omega t} + {\bf E}
e^{-i\omega t}) = \phi_+ e^{-i\omega t} + \phi_-
e^{i\omega t},
\label{field}
\end{equation}
where $\phi_+ = [\phi_-]^* =- e{\bf Er}/2$ and $\omega>0$. We first
consider linearly polarized microwaves whose polarization vector ${\bf
  E}=E\hat{\bf x}$ points along the $x$-direction, i.e. parallel to
the direction of modulation. Results for more general polarizations
will be quoted without derivation.

In the absence of disorder, the Landau level states $|nk \rangle$ in
the Landau gauge remain good eigenstates if the amplitude of the
periodic modulation is small compared to the LL spacing
$\hbar\omega_c$.  ($n$ denotes the LL index and $k$ the momentum in
$y$ direction.)  The corresponding eigenenergies are approximately
given by $\epsilon^0_{nk} \simeq \hbar\omega_c(n+\frac{1}{2}) +
V_{n}\cos(Qk\ell_B^2)$, where the modulation amplitude $V_n$ is given
by $ V_n \simeq  V J_0(qR_c)$ for large LL index $n$.
($\ell_B=(\hbar/eB)^{1/2}$ is the magnetic length, $R_c$ the cyclotron radius,
and $J_n(z)$ denotes a Bessel function.)
\begin{figure}
\psfrag{1}[][][1]{a}
\psfrag{2}[][][1]{b}
\psfrag{3}[][][1]{c}
\psfrag{4}[][][1]{d}
\psfrag{5}[][][1]{e}
\psfrag{6}[][][1]{f}
\includegraphics[width=3.37in]{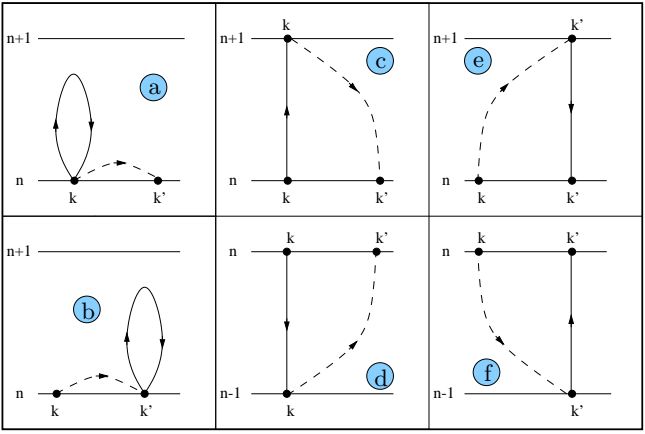}
\caption{
  Relevant processes in the regime $\omega\ll\omega_c$. Full lines
  represent microwave absorption ($\phi_+$) or emission ($\phi_-$) and
  dashed lines disorder scattering (U).  In processes (a) and (b), the
  intermediate states are in the same (valence) LL as the initial and
  final states. As shown in the text, these processes dominate the
  photocurrent. For processes (c)-(f), the LL index of the
  intermediate states differs by one from the LL index of initial and
  final states. Their contribution are smaller by a factor
  $\omega/\omega_c$ than (a) and (b).}
\label{procs}
\end{figure}

To identify the relevant microwave-induced processes, we 
discuss the matrix elements of $U$ and $\phi$. For microwaves
linearly polarized in the $x$ direction, the matrix elements for
absorption and emission are the same and given by
\begin{eqnarray}
\label{matphot}
 \langle n^\prime k' | \phi_\pm | nk \rangle&=&-\frac{eE}{2}k\ell_B^2\delta_{n,n^\prime}\delta_{k,k^\prime}
 \nonumber\\
 &&+\frac{eER_c}{4}\left(\delta_{n,n^\prime- 1}+\delta_{n,n^\prime+ 1}\right)\delta_{k,k^\prime}. 
\end{eqnarray}
Thus, microwaves leave the
LL index unchanged or couple neighboring LLs. By contrast, the
disorder potential has nonzero matrix elements between arbitrary LLs,
\begin{eqnarray}
\label{disordermatrixelement}
 &&  |\langle n^\prime k' | U | nk \rangle|^2
  \nonumber\\
  &&
  \,\,\,\,\,\,\simeq 
\int \frac{d^2q}{(2\pi)^2}
       \delta_{q_y,k'-k} 
\left[J_{\vert n'-n\vert}(qR_c)\right]^2\tilde W({\bf q}).
\end{eqnarray}
This expression is valid in
the limit of large LL indices $n,n'\gg 1$. The dominant microwave-induced
processes contributing to the photocurrent arise from the
contributions
\begin{equation}
    T_\pm=UG_0\phi_\pm +\phi_\pm G_0U 
\end{equation}
to the general $T$-matrix
\begin{equation}
T=U+\phi+ (U+\phi)G_0(U+\phi)+...
\end{equation}
of the system.\cite{Dietel04} Here, $G_0$ denotes the retarded Green
function of the unperturbed system ($U=\phi=0$). We refer to $T_+$
($T_-$) as disorder-assisted microwave absorption (emission).  $T_+$
and $T_-$ can be considered separately, since they contribute
incoherently. The matrix elements of $T_-$ can be shown to equal those
of $T_+$ up to a phase.

We now specialize to the regime $\omega \ll \omega_c$ for
well-separated LLs, which is the focus of this paper. In this case,
only scattering processes with initial and final state in the same LL
are relevant. Disorder-assisted microwave absorption and emission then
proceeds via intermediate states either in the same LL (with amplitude
$M_0$) or in neighboring LLs (with amplitude $M_1$), so that $\langle
n k^\prime\left| T_+ \right|n k \rangle=M_0+M_1$. These processes are
depicted in Fig.\ \ref{procs} (a),(b) and (c)-(f), respectively. We
find that the amplitude $M_1$ is smaller than $M_0$ in the parameter
$\omega/\omega_c$. Thus, we first turn to the contribution $M_0$
which, using Eq.\ (\ref{matphot}), is equal to
\begin{eqnarray}
M_0&=&\langle n k^\prime\left| U \right|n k \rangle G_{0,nk}(\epsilon_{nk}+\omega)\langle n k\left|\phi_+\right|n k \rangle
\nonumber\\
&&\,\,\,\,+\langle n k^\prime\left|\phi_+\right|n k^\prime \rangle G_{0,nk^\prime}(\epsilon_{nk})\langle n k^\prime\left| U \right|n k \rangle \nonumber\\
&=&\frac{eE}{2\omega}(k^\prime-k)\ell_B^2\langle n k^\prime\left| U \right|n k \rangle \quad.
\label{matel}
\end{eqnarray}
Here we used that $G_{0,nk}(\epsilon_{nk}+\omega)=(\epsilon_{nk}+\omega-\epsilon_{nk})^{-1}=\omega^{-1}$ and $ G_{0,nk^\prime}(\epsilon_{nk})=(\epsilon_{nk}-\epsilon_{nk^\prime})^{-1}=-\omega^{-1}$.
With Eq.\ (\ref{disordermatrixelement}), we obtain
\begin{equation}
\label{intermediateintra}
\left|M_0\right|^2
\simeq\left(\frac{eE}{2\omega}\right)^2\int \frac{d^2q}{(2\pi)^2}\delta_{q_y,k^\prime-k} \left[q_y\ell_B^2 J_0(qR_c)\right]^2\tilde{W}({\bf q}),
\end{equation}
valid in the limit of high Landau levels. 

We now turn to an estimate of the contribution $M_1$. The processes
depicted in Fig.\ \ref{procs} (c)-(f) lead to
\begin{widetext}
\begin{eqnarray}
M_1&=&\langle n k^\prime\left| U \right|n+1 k \rangle G_{0,n+1k}(\epsilon_{nk}+\omega)\langle n+1 k\left|\phi_+\right|n k \rangle+ \langle n k^\prime\left| U \right|n-1 k \rangle G_{0,n-1k}(\epsilon_{nk}+\omega)\langle n-1 k\left|\phi_+\right|n k \rangle\nonumber\\
&&+ \langle n k^\prime\left|\phi_+\right|n-1 k^\prime \rangle G_{0,n-1k^\prime}(\epsilon_{nk})\langle n-1 k^\prime\left| U \right|n k \rangle +  \langle n k^\prime\left|\phi_+\right|n+1 k^\prime \rangle G_{0,n+1k^\prime}(\epsilon_{nk})\langle n+1 k^\prime\left| U \right|n k \rangle\nonumber\\
&=&\frac{eER_c}{4}\left[\frac{\langle n k^\prime\left| U \right|n+1 k \rangle}{\omega-\omega_c} +\frac{\langle n k^\prime\left| U \right|n-1 k \rangle}{\omega+\omega_c} - \frac{\langle n-1 k^\prime\left| U \right|n k \rangle}{\omega-\omega_c}-  \frac{\langle n+1 k^\prime\left| U \right|n k \rangle}{\omega+\omega_c}\right]\quad.
\label{M1}
\end{eqnarray}
\end{widetext}
At first sight, the ratio $M_1/M_0$ is of order
$(R_c/q\ell_B^2)(\omega/\omega_c)$. For smooth disorder, $q\sim
1/\xi$, so that $M_1/M_0 \sim (k_F\xi)(\omega/\omega_c)$, where
$k_F\xi\gg 1$. This would imply that $M_1$ can actually dominate over
$M_0$. However, this estimate turns out to be too simplistic. The
reason is that for $\omega\ll \omega_c$, we can write Eq.\ (\ref{M1})
as
\begin{eqnarray}
M_1&=&\frac{eER_c}{4\omega_c}\left\{ [\langle n-1 k^\prime\left| U \right|n k \rangle-\langle n k^\prime\left| U \right|n+1 k \rangle] \right.
\nonumber\\
&&+\left.[ \langle n k^\prime\left| U \right|n-1 k \rangle - \langle n+1 k^\prime\left| U \right|n k \rangle] \right\}\quad.
\end{eqnarray}
We observe that the square brackets involve differences of matrix
elements which differ by a uniform shift by one Landau level. This
leads to a partial cancellation which reduces our previous estimate of
$M_1$ by $q/k_F\sim 1/k_F\xi$. As a result, we find that $M_1/M_0 \sim
\omega/\omega_c$ as claimed above.

\section{Photocurrent}
\label{calc}
\subsection{Mechanisms}

In perturbation theory, disorder-assisted microwave absorption and
emission leads to two contributions to the photocurrent. First, it
changes the electron momentum from $k$ to $k'$, which effectively
corresponds to real-space jumps in the $x$ direction of length
$(k'-k)\ell_B^2$. Due to the applied dc electric field, these jumps
occur preferentially in a fixed direction.  Generalizing the approach
of Titeica \cite{Titeica} to the present situation, this displacement
contribution to the longitudinal photocurrent can be expressed
as\cite{Dietel04}
\begin{widetext}
\begin{equation}
\label{jxphotoI}  
j_x ^{\rm{DP}}=\frac{\pi e}{L_xL_y}\sum_{\sigma=\pm}\sum_{n}\sum_{k,k^\prime}(k^\prime-k)\ell_B ^2 
\left| \langle n k^\prime\left| T_\sigma \right|n k \rangle\right|^2\left[f_{nk} ^0-f_{n k^\prime} ^0\right]\delta(\epsilon_{nk}-\epsilon_{n k^\prime}+\sigma \omega).
\end{equation}
Here $f_{nk} ^0$ is the equilibrium electron distribution function and
$ \epsilon_{nk} = \epsilon_{nk}^0 -eE_{dc}k\ell_B^2$ is the Landau
level energy including the effect of the dc electric field.

Secondly, the microwaves change the electronic distribution function
away from equilibrium. The resulting distribution function
contribution to the longitudinal photocurrent is\cite{Dietel04}
\begin{equation}
j_x^{\rm{DF}}=\frac{\pi e}{L_xL_y}\sum_{n}\sum_{k,k^\prime}(k^\prime-k)\ell_B ^2 \left| \langle n k^\prime\left| U\right|n k \rangle\right|^2\left[\delta f_{nk}-\delta f_{n k^\prime}\right]\delta(\epsilon_{nk}-\epsilon_{n k^\prime}),
\label{jxphotoII}
\end{equation}
\end{widetext}
where $\delta f_{nk}=f_{nk}-f_{nk}^0$ is the deviation of the
nonequilibrium electron distribution function $f_{nk}$ from the
equilibrium distribution $f_{nk}^0$. The distribution function
$f_{nk}$ can be obtained from the kinetic equation
\begin{eqnarray}
   {\frac{\partial f_{nk}}{\partial t}} = \left( {\frac{\partial f_{nk}}{\partial t}} \right)_{\rm dis}
       + \left( {\frac{\partial f_{nk}}{\partial t}} \right)_{\rm mw} - \frac{ f_{\rm nk} - f^0_{\rm nk}}
       {\tau_{\rm in}}.
 \label{kinetic}
\end{eqnarray}
This kinetic equation includes collision integrals for disorder
scattering, $\left( {{\partial f_{nk}}/{\partial t}} \right)_{\rm dis}
=\sum_{n'k'} 2\pi |\langle n'k' | U | nk \rangle |^2 [ f_{n'k'} -
f_{nk} ]\delta ( \epsilon_{nk} - \epsilon_{n'k'} )$, and for
disorder-assisted microwave absorption and emission, $\left(
  {{\partial f_{nk}}/{\partial t}} \right)_{\rm mw}
=\sum_{n'k'}\sum_{\sigma=\pm} 2\pi |\langle n'k' | T_\sigma | nk
\rangle |^2 [ f_{n'k'} - f_{nk} ] \delta ( \epsilon_{nk} -
\epsilon_{n'k'} +\sigma\omega )$. These collision integrals involve
the electron energies including the effects of the dc electric field.
Finally, we include inelastic relaxation within the relaxation-time
approximation, with a phenomenological relaxation time $\tau_{\rm
  in}$.

\subsection{Longitudinal Photocurrent}
\label{long}

In this section, we compute the photocurrent for dc electric fields
applied along the modulation direction.  We first turn to the
distribution-function mechanism which gives the dominant contribution
in the (experimentally relevant) limit of slow inelastic relaxation.
The microwave-induced change in the distribution function, as obtained
from the kinetic equation (\ref{kinetic}), equals
\begin{eqnarray}
   \delta f_{Nk} &=&  \tau_{\rm in} \sum_{k'} \sum_\sigma 2\pi |\langle N k'| T_\sigma |N k\rangle|^2
  \nonumber\\
   &&\hspace*{.3cm}\times(f_{Nk'}^0-f_{Nk}^0)\delta(\epsilon^0_{Nk} - \epsilon^0_{Nk'}  + \sigma \omega)  , 
\label{deltaf}
\end{eqnarray}
where $N$ denotes the valence Landau level in which the Fermi energy
is situated. $\delta f_{nk}$ vanishes for all other Landau levels
$n\neq N$. In the limit of $\omega_c\gg T\gg V$, the distribution
function changes only weakly within the Landau level. Exploiting the
$\delta$ function, we can thus write $f_{Nk'}^0-f_{Nk}^0 \simeq
-\sigma\beta \omega n_F(\epsilon^0_{Nk})[1-n_F(\epsilon^0_{Nk})]$,
where $\beta = 1/k_BT$ and $n_F(\epsilon)$ denotes the Fermi-Dirac
distribution. Noting that to leading order $n_F(\epsilon^0_{Nk})$ is
just the partial filling factor $\nu_N^*$ of the valence Landau level,
we obtain the relation
\begin{equation}
   f_{Nk'}^0-f_{Nk}^0 \simeq -\sigma\beta \omega \nu_N^*(1-\nu_N^*).
\end{equation}
Thus, the change in the distribution function is maximal at the center
of the Landau level and falls off to zero towards the Landau-level
edges. Inserting the expression (\ref{intermediateintra}) for the
matrix element and performing the sum over $k'$, we obtain
\begin{widetext}
\begin{eqnarray}
   \delta f_{Nk} = - 2\pi \tau_{\rm in} \beta \omega \nu_N^*(1-\nu_N^*) \left( eE\over 2\omega\right)^2
    \sum_\sigma \sigma 
\int {d{\bf q}\over (2\pi)^2} [q_y\ell_B^2 J_0(qR_c)]^2\tilde W({\bf q})
         \delta(\epsilon_{Nk}^0-\epsilon_{Nk+q_y}^0 +\sigma\omega).
\end{eqnarray}
\end{widetext}
The ${\bf q}$ integration simplifies significantly in the limit
$\lambda_F \ll a \ll \ell_B^2/\xi$, where it factorizes into an
average over the $\delta$ function and an integral over the remaining
integrand.  The average over the $\delta$ function can be expressed
through the Landau-level density of states,
\begin{eqnarray}
   &&\langle \delta(\epsilon_{Nk}^0-\epsilon_{Nk+q_y}^0 +\sigma\omega)\rangle_k
      \nonumber\\ 
     &&\,\,\,\,\,\,\,\,\,\,\, 
     = 2\pi\ell_B^2 \nu^*(\epsilon_{Nk}^0 +\sigma\omega)\theta(V-|\epsilon_{Nk}^0 +\sigma\omega|).
\end{eqnarray}
Here, we introduced the LL density of states
$\nu^*(\epsilon)=\nu^*\tilde\nu^*(\epsilon)$ where
$\nu^*=(1/2\pi\ell_B^2)(1/\pi V_N)$ denotes the DOS at the LL center
and $\tilde\nu^*(\epsilon)=1/[1-[(\epsilon-E_N)/V_N]^2]^{1/2}$ a
normalized density of states.  Exploiting the fact that the integrand
of the remaining $q$ integration is cut off by the correlator $\tilde
W(q)$ at large $q$, we can replace the Bessel function by an
asymptotic expression for large argument. In this way, we can relate
the integral to the zero-field transport mean free path, defined by
\begin{equation}
  {1\over\tau_{\rm tr}} = {1\over \pi v_F}\int_0^\infty dq (q^2/2k_F^2)\tilde W(q).
\end{equation}
This yields for the change in the distribution function
\begin{eqnarray}
   &&\delta f_{Nk} = - \beta \omega \nu_N^*(1-\nu_N^*) \left( eER_c\over 2\omega\right)^2
   \nonumber\\
     &&\,\,\,\,\,\,\,
     \times\sum_\sigma \sigma {\tau_{\rm in}\over \tau_{\rm tr}^*(\epsilon_{Nk}^0 +\sigma\omega)}
     \theta(V-|\epsilon_{Nk}^0 +\sigma\omega|).
\end{eqnarray}
This expression should be inserted into Eq.\ (\ref{jxphotoII}). Here,
we introduced the transport mean free path $\tau^*_{\rm
  tr}(\epsilon)=\tau_{\rm tr} \nu/\nu^*(\epsilon)$ in the presence of
the magnetic field.  ($\nu$ is the DOS in the absence of the
$B$-field.) In a smooth random potential, this should be distinguished
from the single-particle mean free paths $\tau_s$ and
$\tau_s^*(\epsilon)=\tau_s \nu/\nu^*(\epsilon)$ in the absence and
presence of $B$, respectively.

By a sequence of steps very similar to those for the evaluation of
$\delta f_{Nk}$ just described, we can rewrite Eq.\ (\ref{jxphotoII})
as
\begin{equation}
   \sigma_{xx}^{\rm DF} = \left[e^2 {R_c^2\over 2\tau_{\rm tr}^*} \nu^* \right]
      {2\pi\ell_B^2\over L_xL_y} 2\pi V \sum_k \left(-{\partial \delta f_{Nk}\over \partial\epsilon_{Nk}^0}
      \right) \tilde\nu^*(\epsilon^0_{Nk}).
\end{equation}
Finally performing the sum over $k$, we obtain the result
\begin{eqnarray}
   \sigma_{xx}^{\rm DF} &=& -2 \, \beta \omega \, \nu_N^*(1-\nu_N^*) \left[e^2 {R_c^2\over 2\tau_{\rm tr}^*} \nu^*                \right]\left( eER_c\over 2\omega\right)^2
       {\tau_{\rm in}\over \tau_{\rm tr}^*} 
       \nonumber\\
       &&\times\,\, B_1\left({\omega}/{2V_N}\right),
       \label{DFlongitudinal}
\end{eqnarray}
where 
\begin{equation}
B_1({\omega}/{2V_N})=-{\partial\over\partial \omega} \int_{-V}^{V-\omega} d\epsilon [\tilde\nu^*(\epsilon)]^2
        \tilde\nu^*(\epsilon +\omega)    \,. 
\end{equation}
We obtain for this integral 
\begin{equation}
B_1(x)
=\frac{1}{16}\frac{1-2 x}{\left( x- x^2\right)^{3/2}}\ln \left(\frac{V_N}
    {\Delta}\right).
\end{equation}
Here, $\Delta$ denotes an effective broadening in energy of the LL
edge due to disorder or the dc electric field, both of which cut off
the logarithmic divergence of the integral.

\begin{figure}[t]
\psfrag{longlegendA1}[][][0.6]{$-xA_1(x)$}
\psfrag{longlegendB1}[][][0.6]{$-xB_1(x)/\ln(V_N/\Delta)$}
\psfrag{longA2}[][][0.6]{$-xA_2(x)$}
\psfrag{longB2}[][][0.6]{$-xB_2(x)$}
\psfrag{x}[][][1]{$x$}
\includegraphics[width=3.3in]{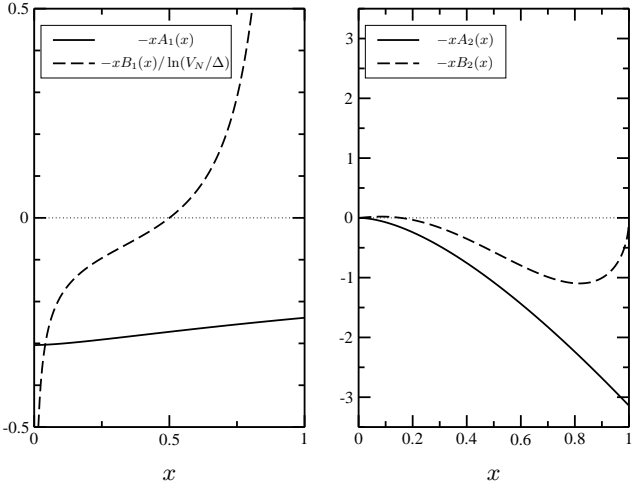}
\caption{Sign and frequency dependence of the photocurrent.
Left panel: functions $-x
A_1(x)$(full line) and $-x
B_1(x)/\ln(V_N/\Delta)$(dashed) determining the sign and frequency
dependence of the displacement and distribution-function contributions
to the longitudinal photoconductivities,
respectively (with $x=\omega/2V_N$).
Right panel: the corresponding functions
$-x  A_2(x)$(full line) and
$-x B_2(x)$(dashed) for the transverse photocurrent. 
}
\label{A1A2}
\end{figure}

The sign and frequency dependence of the longitudinal distribution
function contribution are therefore determined by the function
$-xB_1(x)$. This function is plotted in Fig.\ \ref{A1A2}. We find a
negative photoconductivity in the frequency range $\omega<V_N$. For
larger frequencies $V_N<\omega<2V_N$, the sign of the
photoconductivity changes. This sign change is a specific feature of
our model, arising from the singular density of states at the band
edge for the static periodic modulation potential. While this sign
change is an interesting feature of our model and may be helpful in
distinguishing between the displacement and the distribution function
mechanism in an appropriate experiment, it is not expected to occur in
a more generic situation without a singularity at the Landau level
edge.  Specifically, one expects a negative photoconductivity for all
$\omega<2V_N$ for the case of disorder-broadened Landau levels,
relevant to current experiments.

By a similar calculation, starting from Eq.\ (\ref{jxphotoI}), we obtain 
\begin{eqnarray}
   \sigma_{xx}^{\rm DP} &\propto & -\beta \omega \, \nu_N^*(1-\nu_N^*) \left[e^2 {R_c^2\over 2\tau_{\rm tr}^*} \nu^*                \right]\left( eER_c\over 2\omega\right)^2
       {\tau_{\rm s}\over \tau_{\rm tr}^*} 
       \nonumber\\
       &&\times \,\,A_1\left({\omega}/{2V_N}\right)
\label{sigmaphotoIxx}
\end{eqnarray}
for the longitudinal displacement photoconductivity. We note in
passing that the precise numerical prefactor of $\sigma_{xx}^{\rm DP}$
depends on the details of the smooth-disorder model.  Clearly, this
result is parametrically smaller than the distribution-function
mechanism by a factor $\tau_s^*/\tau_{\rm in}$, where $\tau_s^*$
denotes the single-particle scattering time in the presence of the
magnetic field.  The function $A_1$ appearing in Eq.\ 
(\ref{sigmaphotoIxx}) is given by
\begin{eqnarray}
A_1(x)&=& -  \frac{3}{2\pi}{\partial\over \partial x}    \frac{\ell_B^2}{a} 
\int_{-a/2 \ell_B^2}^{a/2\ell_B^2} 
d q_y  {1\over \sqrt{   
       \sin^2(Qq_y\ell_B^2/2)-x^2       
       }  } 
       \nonumber\\
&=&-\frac{3}{\pi^2}\frac{\partial}{\partial x}K\left(\sqrt{1-x^2}\right)
\end{eqnarray}
and plotted in Fig.\ \ref{A1A2}. Here, $K(x)$ denotes a complete
elliptic function.

The dominant distribution-function contribution to the longitudinal
photocurrent is independent of the microwave polarization. This is
different from the cyclotron-resonance case, \cite{Dietel04} where the
photocurrent depends on the type of circular polarization.
\footnote{The displacement photocurrent
does exhibit a polarization
  dependence, but since this contribution is subdominant for the
  longitudinal photocurrent, we refrain from giving detailed results.}

\subsection{Transverse Photocurrent}
\label{trans}

It was shown in Ref.\ \onlinecite{Dietel04} that within our model, the
contributions of the displacement and the distribution-function
mechanisms to the transverse photocurrent can be of the same order of
magnitude for the cyclotron resonance $\omega\simeq \omega_c$. We find
the same conclusion to hold for intra-LL transitions.

The methods used in Ref.\ \onlinecite{Dietel04} can be readily
extended to intra-LL transitions.  The essential new ingredient in
computing the distribution-function contribution is a drift term
$-eE_{dc}(\partial f_{nk}/\partial k)$ which enters the RHS of the
kinetic equation, Eq.\ (\ref{kinetic}). In this way, we arrive at the
result
\begin{eqnarray}
\sigma_{yy}^{\rm{DF}}&=& -4 \,\beta\omega \, \nu_N^*(1-\nu_N^*)\left[e^2(v_y^2\tau_{\rm{s}}^*)\tilde{\nu}^*\right]\left(\frac{eER_c}{2\omega}\right)^2\frac{\tau_{\rm{ in}}}{\tau_{\rm{tr}}^*} \nonumber\\
&&\times\,\,B_2\left({\omega}/{2V_N}\right)
\end{eqnarray}
where $B_2({\omega}/{2V_N})$ is given by
\begin{equation}
B_2({\omega}/{2V_N})= - \frac{\partial}{\partial \omega} \int_{-V}^{V-\omega} d\epsilon 
\frac{1}{[\tilde{\nu}^*(\epsilon)]^2}\tilde{\nu}^*(\epsilon+\omega) \,. 
\end{equation}
We obtain for this integral 
\begin{eqnarray}
B_2(x) = \left[4x\left(\arcsin(1-2x)+\frac{\pi}{2}\right)-4\sqrt{x-x^2}\right].
\end{eqnarray}
Computing the displacement contribution to the transverse
photoconductivity requires one to evaluate transition rates between
quantum states corresponding to the ``meander" equipotential lines in
the presence of both static periodic modulation and dc electric field.
Following the relevant formalism developed in Ref.\ 
\onlinecite{Dietel04}, we obtain
\begin{eqnarray}
\sigma_{yy}^{\rm{DP}}&=& -2 \, \beta\omega \,\nu_N^*(1-\nu_N^*)\left[e^2(v_y^2\tau_{\rm{s}}^*)\tilde{\nu}^*\right]\left(\frac{eER_c}{2\omega}\right)^2\frac{\tau_{\rm{ in}}}{\tau_{\rm{tr}}^*}
\nonumber\\
&& \times A_2(\omega/2V_N)\left(E^*_{\rm dc}\over E_{\rm dc}\right)^2.
\label{transversesigmaDP}
\end{eqnarray}
The frequency dependence of the photocurrent is described by the
function $A_2$ given by
\begin{eqnarray}
A_2(x)&=&     \frac{\pi\ell_B^2}{a} 
\int_{-a/2 \ell_B^2}^{a/2\ell_B^2} 
d q_y  { x \over \sqrt{   
       \sin^2(Qq_y\ell_B^2/2)-x^2       
       }  } 
\nonumber\\
&=&2xK(\sqrt{1-x^2}).
\end{eqnarray}
A plot of $B_2$ and $A_2$ is provided in Fig. \ref{A1A2}.

Note the singular dependence of the displacement contribution to the
transverse photoconductivity on the dc electric field $E_{dc}$. This
singularity is cut off for small dc electric fields by inelastic
processes when $E_{\rm dc} \sim E^*_{\rm dc}$, where $E_{\rm dc}^*=
Ba/2\pi\sqrt{\tau_{\rm in}\tau_s^*}$.\cite{Dietel04} For $E_{\rm
  dc}\ll E_{\rm dc}^*$, the photoconductivity crosses over to Ohmic
behavior, matching with Eq.\ (\ref{transversesigmaDP}) for $E_{\rm dc}
\sim E^*_{\rm dc}$.\cite{Dietel04} This implies that the contributions
by displacement and distribution mechanisms are of the same order of
magnitude in the transverse case.

We finally remark that both displacement and distribution-function
contribution to the intra-LL transverse photocurrent are independent
of the type of polarization.

\subsection{Comparison with experiment}
\label{exp}

Strictly speaking, our model is different from the experimental
system, due to the assumption of a static periodic potential. However,
previous work\cite{Dietel04} shows that the magnitude of the 
longitudinal photocurrent obtained within our model is 
parametrically identical to that for disorder-broadened 
Landau levels. Specific
features arise within our model due to its anomalous density of states at
the LL edge which leads to additional sign changes of the
photocurrent.

Keeping these caveats in mind, we compare our results to the 
experiment of Ref.\ \onlinecite{Dorozhkin04}. 
Our main results relevant to experiment are:

\begin{itemize} 
\item[(i)] When ignoring effects of the singular density of states at
  the LL edge, the sign of the photocurrent due to intra-LL
  transitions is negative, leading to a reduction of the
  experimentally observed resistivity.
  
\item[(ii)] Comparing the longitudinal photoconductivity in Eq.\ 
  (\ref{DFlongitudinal}) to the dark conductivity $\sigma_{xx}^{\rm
    dark}=e^2(R_c^2/2\tau_{\rm tr}^*)\nu^*$, we find that their ratio
  depends on magnetic field as $\sigma_{xx}^{\rm DF}/\sigma_{xx}^{\rm
    dark} \sim R_c^2/\tau^*_{\rm tr} \sim 1/B$ at fixed $\omega$. This
  magnetic-field dependence actually also holds for inter-LL
  processes.\cite{Dietel04}
  
\item[(iii)] The amplitude of the photoconductivity due to intra-LL
  transitions scales as $\sim 1/\omega$ with the microwave frequency,
  see Eq.\ (\ref{DFlongitudinal}).
  
\item[(iv)] Due to the factor $\nu_N^*(1-\nu_N^*)$, the magnitude of
  the effect is strongest in the LL center and falls off to zero
  towards the LL edge. We note that this filling-factor dependence is
  specific to intra-LL transitions and does not occur for inter-LL
  transitions near the cyclotron resonance or its
  harmonics.\cite{Dietel04}
\end{itemize}

These results are in good agreement with the central experimental
observations. (i) explains the sign of the effect. (ii) is in
agreement with the observations that over the magnetic-field range
where intra-LL processes dominate, the relative microwave-induced
suppression of the conductivity decreases as the magnetic field
increases (see Fig.\ 1 of Ref.\ \onlinecite{Dorozhkin04}).  In
addition, this magnetic-field scaling explains why zero-resistance
states could not be reached in the regime of intra-LL transitions
which occur at higher magnetic fields compared to the cyclotron
resonance or its harmonics. (iii) is in accordance with the
observation that the microwave-induced reduction of the diagonal
resistivity decreases with increasing microwave frequency (see Fig.\ 2
of Ref.\ \onlinecite{Dorozhkin04}).  Finally, (iv) implies that the
photoconductivity suppresses the Shubnikov-deHaas oscillations, an
effect which was very pronounced experimentally.

\section{Summary}
\label{concl}
We have studied the microwave photoconductivity of a 2DEG in a
perpendicular magnetic field with additional unidirectional static
periodic modulation in the regime of intra-LL transitions. We identify
the dominant disorder-assisted microwave absorption and emission
processes for this regime and compute both the longitudinal and
transverse photocurrents.

We find that the distribution-function mechanism dominates for the
longitudinal photocurrent while both distribution-function and
displacement mechanism contribute to the same order to the transverse
photocurrent.  Except for subdominant contributions, we find that the
photoconductivity due to intra-LL processes is polarization
independent. The singular density of states of our model near the LL
edges leads to interesting sign changes, making the photoconductivity
positive in certain frequency ranges. With the exception of these
model-specific predictions, our results are in good agreement with
experiment.  Specifically, we can explain the microwave-induced
suppression of the Shubnikov-deHaas oscillations in the regime of
intra-LL transitions.

{\it Note:} During the completion of this manuscript, we became aware
of related work (V.\ Ryzhii, preprint: cond-mat/0411370; X.L.\ Lei,
S.Y.\ Liu, preprint: cond-mat/0411717) focusing on intra-LL transitions
within different scenarios for the photoconductivity.

\acknowledgments This work has been supported by the DFG-Schwerpunkt
Quanten-Hall-Systeme, and the Junge Akademie.

\begin{appendix}
\end{appendix}
 
\end{document}